\documentclass[preprint,prd,amsmath,showpacs,showkeys,nofootinbib]{revtex4}
\usepackage{graphicx}
\begin{document}
\preprint{IFIC--00--86\mbox{  }FTUV--00--1221}
\title{Hamiltonian lattice QCD at finite density:\\
equation of state in the strong coupling limit}
\author{Yasuo Umino}

\affiliation{
ECT$^*$\\
Strada delle Tabarelle 286\\
I--38050 Villazzano (Trento), Italy\\
and\\
Departamento de F\'{\i}sica Te\`orica, Universitat de Val\`encia \\
E--46100 Burjassot, Val\`encia, Spain}
\date{\today}
\begin{abstract}
The equation of state of Hamiltonian lattice QCD at finite density is examined 
in the strong coupling limit by constructing a solution to the equation 
of motion corresponding to an effective Hamiltonian describing the ground state 
of the many body system. This solution exactly diagonalizes the Hamiltonian to 
second order in field operators for all densities and is used to evaluate the vacuum 
energy density from which we obtain the equation of state. We find that up to and beyond
the chiral symmetry restoration density the pressure of the quark Fermi sea can be negative 
indicating its mechanical instability. Our result is in qualitative agreement 
with continuum models and should be verifiable by future lattice simulations of 
strongly coupled QCD at finite density.
\end{abstract}


\maketitle
\section{\label{Section1}Introduction}
Lattice gauge theory is currently the only known method of solving Quantum Chromodynamics 
(QCD) from first principles. It has developed sufficiently to be able to calculate a broad 
range of low and intermediate energy hadronic observables from ground state hadron masses 
to pion--nucleon scattering lengths. In addition, lattice studies of QCD at finite temperature 
($T$), especially its spectacular success in demonstrating the deconfinement phase transition, 
have been invaluable and continue to play an important role in the search for the 
quark--gluon plasma. 
 
However, as is well known, one of the outstanding problems in lattice gauge theory is the 
consistent implimentation of chemical potential in numerical simulations \cite{cre00}. 
Progress in lattice QCD calculations at finite chemical potential ($\mu$) with 
dynamical quarks has been hindered by the presence of the complex fermion determinant 
which renders standard Monte Carlo techniques useless. In fact, currently there is only 
one numerical method of simulating finite density QCD with three colors at zero temperature. 
This method is based on 
the Monomer--Dimer--Polymer algorithm developed by Karsch and M\"{u}tter \cite{kar89}. 
However its applicability is limited to the strong coupling limit, and furthermore a recent 
study \cite{alo00} indicates that this algorithm might not be reliable for studying the 
chiral phase transition at finite density. Therefore even a qualitative description of 
finite density lattice QCD is welcome. 
  
One method of studying finite density QCD on the lattice is to invoke the strong coupling 
approximation where analytical methods are applicable. Although far from the realistic 
continuum limit, the strong coupling approximation has played an important role in 
the development of
QCD lattice gauge theory from its very inception. In the renowned paper by
Wilson \cite{wil74} this approximation was invoked to demonstrate quark
confinement on the Euclidean space--time lattice. Soon thereafter Kogut and
Susskind \cite{kog75} formulated the Hamiltonian lattice gauge
theory and concluded that in the strong coupling limit the quark dynamics is
best described by a collection of non--Abelian electric flux tubes with quarks
attached at their ends. This was followed by the work of Baluni and
Willemsen who used a variant of the Kogut--Susskind formalism to demonstrate 
quantitatively that dynamical chiral symmetry breaking indeed takes places in 
lattice QCD at strong coupling \cite{bal76}. Finally, calculations by Kogut, 
Pearson and Shigemitsu \cite{kog79} and by Creutz \cite{cre79} suggesting the absence 
of a phase transition between the strong and weak coupling regimes of QCD
motivated numerous studies using the strong coupling approximation. 

Strong coupling QCD at finite $T$ and/or $\mu$ has previously been studied 
analytically both in the Euclidean \cite{dam85,ilg85,bil92} and in the 
Hamiltonian \cite{pat84,van84,ley88,gre00} formulations. One of the main objectives of 
these studies was to investigate the nature of chiral phase transition at 
finite temperature and density. In each study this was accomplished by constructing some 
effective action or Hamiltonian for strongly coupled lattice QCD using Kogut--Susskind 
fermions. These effective descriptions involve the introduction of composite meson and 
baryon fields which are treated in the mean field approximation.\footnote{The 
Monomer--Dimer--Polymer algorithm \cite{kar89} also uses composite meson and baryons fields.} 
The consensus is that at zero or low temperatures strong coupling QCD undergoes a first order 
chiral phase transition from the broken symmetry phase below a critical chemical potential 
$\mu_{\text{C}}$ to a chirally symmetric phase above $\mu_{\text{C}}$. The only exception 
is the work by Le Yaouanc et al. \cite{ley88} which does not involve effective composite 
fields but is equivalent to the mean field approximation. In this case the chiral 
phase transition was found to be of second order.

In this paper we present the equation of state of Hamiltonian lattice QCD at finite 
density in the strong coupling limit using both Kogut--Susskind and Wilson fermions. 
As in previous studies we begin 
with an effective theory by using a Hamiltonian describing the ground state of 
strongly coupled QCD. However, our approach differs from earlier works in that we 
do not introduce composite fields but explictly construct a solution to the 
field equations of motion corresponding to the effective Hamiltonian. This solution 
exactly diagonalizes the Hamiltonian to second order in field operators for all densities 
and is used to calculate the vacuum energy density from which we obtain the equation of state. 
We find that up to and beyond the chiral symmetry restoration density the quark Fermi 
sea can have negative pressure indicating its mechanical instability. Our result is in
qualitative agreement with those obtained using continuum effective QCD models 
\cite{bub96,alf98} and should be verifiable by future lattice simulationns of strongly 
coupled QCD at finite density.

Our approach admits to a systematic extension to finite temperature and to the 
description of bound states. In fact, we first introduce temperature 
and chemical potential simultaneously into our formalism and then take the limit of vanishing 
temperature to examine the consequences. Description of bound states is accomplished 
by interpreting our solution within the context of the N--quantum approach (NQA) to 
quantum field theory \cite{gre65,gre94} which we shall discuss in the concluding section.
In the same section we propose how the NQA may be combined with the present approach 
to study the nature of the deconfinement phase transition.

In Section~\ref{Section2} we introduce our effective Hamiltonian for the ground state of 
strong coupling QCD using Wilson fermions, and discuss the condition under which it can be 
extended to finite $T$ and $\mu$. Our ansatz for the lattice quark field at finite $T$ and 
$\mu$ is presented in Section~\ref{Section3}. The equation of motion at finite $\mu$ is then 
derived in Section~\ref{Section4} and used to diagonalize the effective Hamiltonian to 
second order in field operators and to evaluate the vacuum energy density. In the same 
section we determine the unknown quantities in our ansatz by deriving coupled equations 
for the dynamical quark mass and the total chemical potential and solving them 
self--consistently. Having constructed a solution for the quark field we present in 
Section~\ref{Section5} the equation of state of Hamiltonian lattice QCD at finite density 
in the strong coupling approximation. We summarize 
our results in Section~\ref{Section6} and discuss how our approach may be extended to 
incorporate temperature and to describe the deconfinement phase transition. A review of the 
properties of free lattice Wilson fermions using the Hamiltonian formulation is given in the 
Appendix.

\section{\label{Section2}The effective Hamiltonian}

We begin by introducing Smit's effective Hamiltonian \cite{smi80} describing the ground 
state of strongly coupled QCD. This state is the one in which 
no links are excited by the electric flux. It is also infinitely degenerate since various 
color singlet states may be created at each lattice site without increasing
the ground state energy. This degeneracy is lifted by the propagation of quarks on the
lattice. The simplest type of such a propagation involves a quark exciting a flux link
and an anti--quark deexciting the same link and corresponds to the propagation of a 
meson. Smit obtained an effective Hamiltonian describing this propagation using second
order perturbation theory involving only the quark field $\Psi$ with a nearest 
neighbour interaction. The Hamiltonian is effective because it only acts on the 
space of states with no excited links. Nevertheless, it serves our purpose since the 
main quantity of interest in this work is the vacuum energy density which is obtained 
by diagonalizing Smit's Hamiltonian.

In the Hamiltonian formulation of lattice field theory \cite{kog75}
only the spatial coordinates are discreticized while the temporal coordinate remain
continuous. We adopt the notation of Smit \cite{smi80} where the discrete sums over the 
spatial and momentum coordinates are given by
\begin{equation}
\sum_{\vec{x}} = a^3 \sum_{\vec{m}}
\end{equation}
and 
\begin{equation}
\sum_{\vec{p}} = \frac{1}{V} \sum_{\vec {n}}
\end{equation}
with $\vec{x} = a \vec{m} = a (m_1, m_2, m_3)$ and 
$\vec{p} = \frac{\pi}{La}\vec{n} = \frac{\pi}{La} (n_1, n_2, n_3)$. Here $a$ is the 
lattice spacing and $L$ defines the number of unit lattice cells with 
$m_l, n_l = 0, \pm 1, \pm 2, ..., \pm L$. With this notation the volume $V$ is given
by $V = (2La)^3$. Henceforth we shall work in lattice units where $a=1$ so that 
$-\pi \leq p_l \leq +\pi$, and use Wilson fermions. 

The effective Hamiltonian derived by Smit \cite{smi80} using the temporal gauge is 
\begin{eqnarray}
H_{\text{eff}}
& = &
M_0 \sum_{\vec{x}} \bigl(\Psi^\dagger_{a\alpha}\bigr)_\rho(\vec{x}\,)
\left( \gamma_0\right)_{\rho\nu} 
\bigl(\Psi_{a\alpha}\bigr)_\nu(\vec{x}\,) 
\nonumber \\
&   & -\frac{K}{2N_{\text{c}}} \sum_{\vec{x}}\sum_l
\nonumber \\
&   &
\otimes \Biggl[
\bigl(\Psi^\dagger_{a\alpha}\bigr)_\rho(\vec{x}\,)  
\left( \Sigma_l \right)_{\rho\nu} 
\bigl(\Psi_{b\alpha}\bigr)_\nu(\vec{x} + \hat{a}_l)
\bigl(\Psi^\dagger_{b\beta}\bigr)_\gamma(\vec{x} + \hat{a}_l)  
\left( \Sigma_l \right)^\dagger_{\gamma\delta} 
\bigl(\Psi_{a\beta}\bigr)_\delta(\vec{x}\,)
\nonumber \\
&   &
\;\;\;\;\;
+ 
\bigl(\Psi^\dagger_{b\beta}\bigr)_\gamma(\vec{x} + \hat{a}_l)  
\left( \Sigma_l \right)^\dagger_{\gamma\delta} 
\bigl(\Psi_{a\beta}\bigr)_\delta(\vec{x}\,)
\bigl(\Psi^\dagger_{a\alpha}\bigr)_\rho(\vec{x}\,)  
\left( \Sigma_l \right)_{\rho\nu} 
\bigl( \Psi_{b\alpha}\bigr)_\nu(\vec{x} + \hat{a}_l)
\Biggr]
\label{Heffx}
\end{eqnarray}
where $\Sigma_l \equiv -i(\gamma_0\gamma_l - ir\gamma_0)$ and $\hat{a}_l$ is a unit vector 
along the positive $l$--axis. We denote color, flavor and Dirac indices by $(ab)$, 
$(\alpha\beta)$ and $(\rho\nu\gamma\delta)$, respectively. Summation convention for repeated 
indices is implied. The three parameters in this Hamiltonian are the Wilson parameter $r$ 
which takes on values between 0 and 1, the current quark mass $M_0$ and the effective coupling 
constant $K = 2N_{\text{c}}/(N_{\text{c}}^2-1)\,1/g^2$ where $g$ is the QCD coupling constant. 
$N_{\text{c}}$ is the number of colors. When $r = 0$ the quark fields become Kogut--Susskind 
fermions. 

Smit's Hamiltonian is valid to order ${\cal O}\,(1/g^2)$ in the strong coupling expansion. 
The ${\cal O}\,(1/g^2)$ corrections involve products of quark bilinears which describe meson 
propagation mentioned above and are known as ``meson terms''.  For $N_{\text{c}} = 3$, 
contributions from the subsequent order in the $1/g^2$ expansion would consist of products 
of terms which are trilinear in the quark fields called ``baryon terms''. These meson 
and baryon terms appear in the strong coupling expansions of both Euclidean and Hamiltonian 
lattice QCD and are the motivations for introducing effective composite meson and baryon 
fields. In this work we do not take the baryon terms into account but our formalism presented 
here is also applicable if such terms were present in the effective Hamiltonian.

In the absence of the current quark mass and the Wilson parameter ($M_0 = r = 0$), 
$H_{\text{eff}}$ posseses a $U(4N_{\text{f}})$ symmetry with $N_{\text{f}}$ being the 
number of flavors. This symmetry is spontaneously broken to 
$U(2N_{\text{f}})\otimes U(2N_{\text{f}})$ accompanied by the appearance of 
$8N_{\text{f}}^2$ Goldstone bosons \cite{smi80}. A finite current quark mass 
also breaks the original $U(4N_{\text{f}})$ symmetry, albeit explicitly, to 
$U(2N_{\text{f}})\otimes U(2N_{\text{f}})$. Introduction of the Wilson term explicitly 
breaks the latter symmetry further down to $U(N_{\text{f}})$ thereby solving the fermion 
doubling problem. 

We shall work exclusively in momentum space. Our convention for the Fourier transform 
from configuration to momentum space is 
$\Psi(\vec{x}\,) =  \sum_{\vec{p}}\Psi(\vec{p}\,)e^{\vec{p}\cdot\vec{x}}$, which implies 
that the volume $V$ is given by $V = \sum_{\vec{x}} = \delta_{\vec{p},\vec{p}}.$ Then the 
charge conjugation symmetric form of Smit's Hamiltonian in momentum space is given by
\begin{eqnarray}
H_{\text{eff}}
& = &
\frac{1}{2} \sum_{\vec{p}}
M_0 \left( \gamma_0\right)_{\rho\nu}
\biggl[ \bigl(\Psi^\dagger_{a\alpha}\bigr)_\rho(\vec{p}\,),
\bigl(\Psi_{a\alpha}\bigr)_\nu(-\vec{p}\,) \biggr]_-
\nonumber \\
&   & 
-\frac{K}{8N_{\text{c}}} \sum_{\vec{p}_1,\ldots,\vec{p}_4 }\sum_{l}
\delta_{\vec{p}_1+\cdot\cdot\cdot+\vec{p}_4,\vec{0}}
\Biggl[ e^{i((\vec{p}_1+\vec{p}_2)\cdot \hat{n}_l)} 
+ e^{i((\vec{p}_3+\vec{p}_4)\cdot \hat{n}_l)} \Biggr]
\nonumber \\
&   &
\otimes \Biggl[
\bigl(\Psi^\dagger_{a\alpha}\bigr)_\rho(\vec{p}_1\,)  
\left( \Sigma_l \right)_{\rho\nu} 
\bigl( \Psi_{b\alpha}\bigr)_\nu(\vec{p}_2\,)
-
\bigl(\Psi_{a\alpha}\bigr)_\nu(\vec{p}_1\,)
\left( \Sigma_l \right)^\dagger_{\rho\nu}
\bigl( \Psi^\dagger_{b\alpha}\bigr)_\rho(\vec{p}_2\,) \Biggr]
\nonumber \\
&   &
\otimes \Biggl[
\bigl(\Psi^\dagger_{b\beta}\bigr)_\gamma(\vec{p}_3\,)  
\left( \Sigma_l \right)^\dagger_{\gamma\delta} 
\bigl(\Psi_{a\beta}\bigr)_\delta(\vec{p}_4\,)
-
\bigl(\Psi_{b\beta}\bigr)_\delta(\vec{p}_3\,)
\left( \Sigma_l \right)_{\gamma\delta}
\bigl( \Psi^\dagger_{a\beta}\bigr)_\gamma(\vec{p}_4\,) \Biggr]
\label{eq:Heffp}
\end{eqnarray}
This effective Hamiltonian is the starting point of the present investigation. Our method 
for obtaining the equation of state consists of extending $H_{\text{eff}}$ to finite $\mu$ 
and constructing a quark field operator $\Psi$ which diagonalizes the Hamiltonian to second 
order in field operators for all densities. Once this solution has been found it can be used 
to evaluate the vacuum energy density from which we obtain the pressure of the many body system.

However before extending $H_{\text{eff}}$ to finite $T$ and/or $\mu$, it is necessary
to impose a condition on these external parameters so that all links would remain in 
their ground states. In the strong coupling limit the amount of energy required to excite 
one color electric flux link is
\begin{equation}
E = \frac{1}{2N_{\text{c}}}(N_{\text{c}}^2 - 1)\,g^2 = \frac{1}{K}
\label{eq:ELINK}
\end{equation}
Therefore an extension of $H_{\text{eff}}$ to finite $T$ and/or $\mu$ will be valid as long as 
$T,\,\mu < 1/K$ \cite{ley88}\footnote{Note that in \cite{ley88} $E$ has been approximated 
by $E \approx N_{\text{c}}\,g^2$.} since the Hamiltonian only acts on the space of states 
with no excited links. We shall see that this condition is satisfied in the present work.

The effective Hamiltonian is extended to finite $T$ and $\mu$ in two steps. 
The first one is to make the following trivial replacement of the current quark mass term 
in $H_{\text{eff}}$ 
\begin{equation}
M_0 \left( \gamma_0\right)_{\rho\nu} \rightarrow 
M_0 \left( \gamma_0\right)_{\rho\nu} - \mu_0\, \delta_{\rho\nu} 
\end{equation}
where $\mu_0$ is the quark chemical potential. Note that $\mu_0$ should {\em not} be 
identified with the total chemical potential $\mu_{\text{tot}}$ of the interacting many 
body system. As we shall see below, the ${\cal O}\,(1/g^2)$ interaction terms in 
$H_{\text{eff}}$ will induce a correction to $\mu_0$ which in general is 
momentum dependent. We shall therefore refer to $\mu_0$ as the "bare" quark chemical potential 
and treat it as an input parameter. The second step is to introduce an ansatz for the 
quark field at finite $T$ and $\mu$.

\section{\label{Section3}An ansatz for finite $T$ and $\mu$}
We proceed by presenting our ansatz for the $\Psi$ field in $H_{\text{eff}}$ at any  
temperature and chemical potential. The special case of this ansatz for free space was 
given in \cite{umi00a}. It has the same structure as the free lattice Dirac field and obeys 
the free lattice Dirac equation with a mass which is interpreted as the dynamical quark 
mass. This mass is the only unknown quantity in the free space ansatz and is determined 
by solving a gap equation. It was shown in \cite{umi00a} that this ansatz exactly diagonalizes 
$H_{\text{eff}}$ to second order in field operators. Properties of free lattice  Dirac 
fields using Wilson fermions are given in the Appendix.
 
Temporarily dropping color and flavor indices the free space ansatz given in \cite{umi00a} is 
\begin{equation}
\Psi_\nu^{\text{Free}}(t,\vec{p}\,) = 
b(\vec{p}\,) \xi_\nu(\vec{p}\,)e^{-i\omega(\vec{p}\,) t}
+ d^\dagger(-\vec{p}\,)\eta_\nu(-\vec{p}\,)e^{+i\omega(\vec{p}\,) t}
\label{eq:psip}
\end{equation}
with $\nu$ denoting the Dirac index. The annihilation operators for particles $b$ and 
anti--particles $d$ annihilate an interacting vacuum state $|\,{\cal G}_0\,\rangle$, and 
obey the free fermion anti--commutation relations. The properties of the lattice spinors 
$\xi$ and $\eta$ are given in the Appendix. The free lattice Dirac equation fixes the 
excitation energy $\omega(\vec{p}\,)$ to be
\begin{equation}
\omega(\vec{p}\,) = \left( \sum_l {\rm sin}^2 (\vec{p}\cdot\hat{n}_l)
+ M^2(\vec{p}\,) \right)^{1/2}
\label{eq:EXEN}
\end{equation}
where $M(\vec{p}\,)$ is the dynamical quark mass.
 
In order to extend Eq.~(\ref{eq:psip}) to finite $T$ and $\mu$ we observe that the 
annihilation operators $b$ and $d$ in $\Psi^{\text{Free}}$ no longer annihilate the 
interacting vacuum state at finite $T$ and $\mu$ denoted as $|\,{\cal G}(T,\mu)\rangle$. 
To construct operators that annihilate $|\,{\cal G}(T,\mu)\rangle$ we apply a 
generalized thermal Bogoliubov transformation to the $b$ and $d$ operators following 
the formalism of thermal field dynamics \cite{umezawa}
\begin{subequations}
\label{eq:TBT}
\begin{eqnarray}
b(\vec{p}\:) 
& = & 
\alpha_p B(\vec{p}\:) - \beta_p \tilde{B}^{\dagger}(-\vec{p}\:)
\label{equationa} \\
d(\vec{p}\:) 
& = &
\gamma_p D(\vec{p}\:) - \delta_p \tilde{D}^{\dagger}(-\vec{p}\:)
\label{equationb}
\end{eqnarray}
\end{subequations}
The thermal field operators $B$ and $\tilde{B}^\dagger$
annihilate a quasi--particle and  create a quasi--hole at finite $T$ and
$\mu$, respectively, while $D$ and  $\tilde{D}^\dagger$ are the
annihilation operator for a quasi--anti--particle and creation opertor for
a quasi--anti--hole, respectively. 

These thermal annihilation operators
annihilate the interacting thermal vacuum state {\em for each $T$ and $\mu$}.  
\begin{equation}
B(\vec{p}\:) |\,{\cal G}(T,\mu)\rangle 
=\tilde{B}(\vec{p}\:) |\,{\cal G}(T,\mu)\rangle 
= D(\vec{p}\:) |\,{\cal G}(T,\mu)\rangle 
=\tilde{D}(\vec{p}\:) |\,{\cal G}(T,\mu)\rangle 
= 0
\label{eq:AVAC}
\end{equation}
We note that the thermal doubling of the Hilbert space accompanying the thermal
Bogoliubov transformation is implicit in Eq.~(\ref{eq:AVAC})
where a vacuum state which is annihilated by operators 
$B$, $\tilde{B}$, $D$ and $\tilde{D}$ is defined. In addition, since we
shall be working only in the space of quantum field operators it 
is not necessary to specify the structure of the thermal vacuum $|\, {\cal G}(T,\mu)\rangle$. 

The thermal operators also satisfy the Fermion anti--commutation relations
\begin{equation}
\biggl[B^{\dagger}(\vec{p}\:), B(\vec{q}\:) \biggr]_+ =
\biggl[D^{\dagger}(\vec{p}\:), D(\vec{q}\:) \biggr]_+ = 
\biggl[\tilde{B}^{\dagger}(\vec{p}\:), \tilde{B}(\vec{q}\:) \biggr]_+ =
\biggl[\tilde{D}^{\dagger}(\vec{p}\:), \tilde{D}(\vec{q}\:) \biggr]_+ =
\delta_{\vec{p},\vec{q}}
\label{eq:THOP}
\end{equation}
with vanishing anti--commutators for the remaining combinations.
The coefficients of the transformation are 
\begin{subequations}
\label{eq:COEFF}
\begin{eqnarray}
\alpha_p & = & \sqrt{1-n_p^-} \label{eq:COEFFa} \\
\beta_p & = & \sqrt{n_p^-} \label{eq:COEFFb} \\
\gamma_p & = & \sqrt{1-n_p^+} \label{eq:COEFFc} \\
\delta_p & = & \sqrt{n_p^+} \label{eq:COEFFd}
\end{eqnarray}
\end{subequations}
where 
\begin{equation}
n_p^{\pm} = \frac{1}{e^{(\omega(\vec{p}\,) \pm \mu)/T}+1}
\end{equation}
are the Fermi distribution functions for particles ($n_p^-$) and anti--particles ($n_p^+$). 
We stress that the chemical potential appearing in the Fermi distribution functions is 
the {\em total} chemical potential of the interacting many body system. The coefficients 
are chosen so that $n_p^\pm$ are given by
\begin{subequations}
\label{eq:DIST}
\begin{eqnarray}
n_p^- 
& = &
\langle {\cal G}(T,\mu)|b^{\dagger}(\vec{p}\:)b(\vec{p}\:)
|{\cal G}(T,\mu)\rangle\\
n_p^+
& = &
\langle {\cal G}(T,\mu)|d^{\dagger}(\vec{p}\:)d(\vec{p}\:)
|{\cal G}(T,\mu)\rangle %
\end{eqnarray}
\end{subequations}
Hence in this approach temperature and chemical potential are introduced simultaneously 
through the coefficients of the thermal Bogoliubov transformation and are treated on an 
equal footing. 

After applying the Bogoliubov transformation to Eq.~(\ref{eq:psip}) our ansatz at finite $T$ 
and $\mu$ becomes
\begin{eqnarray}
\Psi_\nu(t, \vec{p}\:) 
& = &
\biggl[ \alpha_p B(\vec{p}\:) - \beta_p \tilde{B}^{\dagger}(-\vec{p}\:)\biggr]
\xi_\nu(\vec{p}\,)e^{-i[\omega(\vec{p}\,)-\mu_{\text{tot}}] t}  \nonumber \\
&    &
\;\;\;\;\;\;\;\;\;\;\;\;\;\;\;\;\;\;\;\; +
\biggl[ \gamma_p D^{\dagger}(-\vec{p}\:) - \delta_p
\tilde{D}(\vec{p}\:) \biggr] \eta_\nu(-\vec{p}\,)e^{+i[\omega(\vec{p}\,) +
\mu_{\text{tot}}] t}
\label{eq:HAAGTMU}
\end{eqnarray}
and satisfies the equation of motion corresponding to the free lattice Dirac Hamiltonian 
at finite chemical potential given by
\begin{equation}
H^0 = \frac{1}{2} \sum_{\vec{p}}
\Biggl[ - \sum_l \text{sin}(\vec{p}\cdot\hat{n}_l)(\gamma_0\gamma_l)_{\eta\nu}
+ M(\vec{p}\,)(\gamma_0)_{\eta\nu} - \mu_{\text{tot}} \delta_{\eta\nu} \Biggr]
\Bigg[ \Psi_\eta^\dagger(t, \vec{p}\,), \Psi_\nu(t, \vec{p}\,) \Biggr]_-
\label{eq:H0}
\end{equation}
The spinors $\xi$ and $\eta$ in Eq.~(\ref{eq:HAAGTMU}) obey the same properties as 
in free space and the excitation energy $\omega(\vec{p}\,)$ has the same form as in 
Eq.~(\ref{eq:EXEN}). The unknown quantities in our ansatz Eq.~(\ref{eq:HAAGTMU}) are 
the dynamical quark mass $M(\vec{p}\,)$ and the total chemical potential $\mu_{\text{tot}}$ 
which will be determined in the following section. 

In this work we shall take the $T \rightarrow 0$ limit which amounts to setting 
$\gamma_p = 1$ and $\delta_p = 0$ in the Bogoliubov transformation Eq.~(\ref{eq:TBT}) 
thereby suppressing the excitation of anti--holes. In this limit $\beta_p^2$ becomes 
the Heaviside function $\beta_p^2 = \theta( \mu_{\text{tot}} - \omega(\vec{p}\,))$ 
defining the Fermi momentum $\vec{p}_F$ through the relation  
\begin{equation}
\mu_{\text{tot}} = \left( \sum_l {\rm sin}^2 (\vec{p}_F\cdot\hat{n}_l) 
+ M^2(\vec{p}_F)\right)^{1/2}
\end{equation}
Note that we define chemical potential such that $\mu_{\text{tot}} \geq  M(\vec{p}_F)$ which 
differs from the conventional definition of chemical potential used in lattice calculations 
where $\mu \geq 0$.

One of the simplest quantities to calculate using the ansatz of Eq.~(\ref{eq:HAAGTMU}) in 
the $T \rightarrow 0$ limit is the quark number density $n$ given by
\begin{subequations}
\label{eq:NDEN}
\begin{eqnarray}
n & = & \frac{1}{2 V N_{\text{f}} N_{\text{c}}}\langle\bar{\Psi}\gamma_0\Psi\rangle  
= \frac{1}{2 V N_{\text{f}} N_{\text{c}}}
\frac{1}{2} \sum_{\vec{p}} \langle 
: [\bigl(\bar{\Psi}^\dagger_{a,\alpha}\bigr)_\rho(\vec{p}\,),
\bigl(\Psi_{a,\alpha}\bigr)_\nu(-\vec{p}\,)]_- : \rangle\, (\gamma_0)_{\rho\nu}
\label{eq:NDENa}\\
& = &
\sum_{\vec{p}} \beta^2_p 
= \sum_{\vec{p}} \theta( \mu_{\text{tot}} - \omega(\vec{p}\,)) 
\end{eqnarray}
\end{subequations}
where the symbol :  : denotes normal ordering with respect to the vacuum at zero 
temperature $|\, {\cal G}(T=0,\mu)\rangle$. Therefore, above a sufficiently large value of 
$\mu_{\text{tot}}$ the quark number density becomes a constant which with the present 
normalization will equal unity. This saturation effect is purely a lattice artifact 
originating from the ${\text{sin}}^2 (\vec{p}\cdot\hat{n}_l)$ 
term in $\omega(\vec{p}\,)$. 

Another quantity that may be readily calculated using the $T \rightarrow 0$ ansatz is the 
chiral condensate. It is found to be proprotional to the dynamical quark mass
\begin{equation}
\frac{1}{2V N_{\text{f}} N_{\text{c}} }\langle \bar{\Psi}\Psi \rangle = 
-\sum_{\vec{p}} \alpha_p^2 \frac{M(\vec{p}\,)}{\omega(\vec{p}\,)}
\label{eq:cond}
\end{equation}
Below we shall derive a gap equation for $M(\vec{p}\,)$ and show that for a given 
physically reasonable set of parameters there exists a critical chemical potential 
above which $M(\vec{p}\,) = 0$. Thus the chiral condensate may be identified as being 
the order parameter for the chiral phase transition at finite density.

\section{\label{Section4}Applications of the Equation of Motion}

\subsection{\label{Section4.1}The equation of motion}
We now calculate the equation of motion corresponding to $H_{\text{eff}}$ with
our ansatz for finite $\mu$ using two light flavors. The result is used to show 
that our ansatz exactly diagonalizes the Hamiltonian to second order in field 
operators for all densities and to calculate the vacuum energy density. In addition, 
by analyzing the Dirac structure of the equation of motion we derive coupled equations 
for the dynamical quark mass and the total chemical potential. They are solved to 
lowest order in the $1/N_{\text{c}}$ expansion thereby completing our construction 
of a solution to the lattice field theory defined by $H_{\text{eff}}$.

The equation of motion for $H_{\text{eff}}$ is obtained by exploiting the fact that 
our ansatz also satisfies the equation of motion corresponding to the free lattice Dirac 
Hamiltonian $H^0$ given in Eq.~(\ref{eq:H0}). We therefore have the relation
\begin{equation}
:\Bigl[ \bigl(\Psi_{a\alpha}\bigr)_\nu(t, \vec{q}\,), H_{\text{eff}} \Bigr]_- :
\;\;\;\;
=
\;\;\;\;
:\Bigl[ \bigl(\Psi_{a\alpha}\bigr)_\nu(t, \vec{q}\,), H^0 \Bigr]_- :
\label{eq:CRUX}
\end{equation}
which plays a crucial role in our construction of a solution for the quark field $\Psi$. 
Evaluating both sides of Eq.~(\ref{eq:CRUX}) and equating terms which are linear in the 
field operators we obtain the equation of motion for $\Psi$
\begin{eqnarray}
\lefteqn{ \!\!\!\!\!\!\!\!\!\!\!\!\!\!\!
\Bigl[ \sum_l \text{sin}(\vec{q}\cdot\hat{n}_l)(\gamma_0\gamma_l)_{\rho\delta}
+ M(\vec{q}\,)(\gamma_0)_{\rho\delta} - \mu_{\text{tot}}\delta_{\rho\delta} \Bigr]
\bigl(\Psi_{a\alpha}\bigr)_\delta(t, \vec{q}\,) =
} \nonumber \\
&   &
\Biggl\{
M_0 \left(\gamma_0\right)_{\rho\delta} - \mu_0 \delta_{\rho\delta}
\nonumber \\
&   & 
+ \frac{1}{N_{\text{c}}}K\sum_{\vec{p}}\sum_l 
\alpha_p^2 \Lambda^+_{\nu\gamma}(\vec{p}\,)
\nonumber \\
&   &
\otimes\left[ \cos\left( \vec{p}-\vec{q}\;\right)\cdot\hat{n}_l
\Biggl(
\bigl(\Sigma_l\bigr)_{\gamma\nu}\bigl(\Sigma_l\bigr)^\dagger_{\rho\delta}
+
\bigl(\Sigma_l\bigr)^\dagger_{\rho\nu}\bigl(\Sigma_l\bigr)_{\gamma\delta}
\Biggr)\right.
\nonumber \\
&   &
\:\:\:\:\:\:\:\:\:\:
\left.
+
\cos\left( \vec{p}+\vec{q}\;\right)\cdot\hat{n}_l
\Biggl(
\bigl(\Sigma_l\bigr)^\dagger_{\gamma\nu}\bigl(\Sigma_l\bigr)^\dagger_{\rho\delta}
+
\bigl(\Sigma_l\bigr)_{\rho\nu}\bigl(\Sigma_l\bigr)_{\gamma\delta}
\Biggr)\right]
\nonumber \\
&   &
-\frac{1}{N_{\text{c}}}\frac{K}{4} \sum_{\vec{p}} \sum_l 
\left[ 2\alpha_p^2 \Lambda^+_{\nu\gamma}(\vec{p}\,) - \delta_{\nu\gamma} \right]
\nonumber \\
&   &
\otimes\left[ N_{\text{c}}
\Biggl(
\bigl(\Sigma_l\bigr)_{\rho\nu}\bigl(\Sigma_l\bigr)^\dagger_{\gamma\delta}
+
\bigl(\Sigma_l\bigr)^\dagger_{\rho\nu}\bigl(\Sigma_l\bigr)_{\gamma\delta}
\Biggr)\right.
\nonumber \\
&   &
\:\:\:\:\:\:\:\:\:\:
\left.
+
\cos\left( \vec{p}+\vec{q}\;\right)\cdot\hat{n}_l
\Biggl(
\bigl(\Sigma_l\bigr)^\dagger_{\rho\nu}\bigl(\Sigma_l\bigr)^\dagger_{\gamma\delta}
+
\bigl(\Sigma_l\bigr)_{\rho\nu}\bigl(\Sigma_l\bigr)_{\gamma\delta}
\Biggr)\right] \Biggr\} \bigl(\Psi_{a\alpha}\bigr)_\delta(t, \vec{q}\,) 
\label{eq:EOM1} 
\end{eqnarray}
with $\Lambda^+(\vec{p}\,) \equiv  \xi(\vec{p}\,)\otimes\xi^\dagger(\vec{p}\,)$
being the positive energy projection operator defined in Eq.~(\ref{eq:projp}).

\subsection{\label{Section4.2}Diagonalization of $H_{\text{eff}}$ and the vacuum energy density}
We shall now show that our $T \rightarrow 0$ ansatz exactly diagonalizes the effective 
Hamiltonian to second order in field operators. The diagonalization procedure involves only 
algebraic substitutions and does not require any 
approximations. The quantity of interest here is the off--diagonal Hamiltonian which, to 
second order in field operators, is found to be
\begin{eqnarray}
H_{\text{off}}|\;{\cal G}(0, \mu)\rangle
& = &
- \sum_{\vec{q}} \Biggl\{ \alpha_q \xi^\dagger_\rho(\vec{q}\,)
\left[ M_0\bigl( \gamma_0 \bigr)_{\rho\delta} - \mu_0 \delta_{\rho\delta} \right]
\nonumber \\
&   &
+ \frac{1}{N_{\text{c}}}K\sum_{\vec{p}}\sum_l \alpha_p^2 \,\alpha_q\,
\Lambda^+_{\nu\rho}(\vec{p}\,)
\nonumber \\
&   &
\otimes\; \xi^\dagger_\gamma(\vec{q}\,)
\left[ \cos\left( \vec{p}-\vec{q}\;\right)\cdot\hat{n}_l
\Biggl(
\bigl(\Sigma_l\bigr)_{\rho\nu}\bigl(\Sigma_l\bigr)^\dagger_{\gamma\delta}
+
\bigl(\Sigma_l\bigr)^\dagger_{\rho\nu}\bigl(\Sigma_l\bigr)_{\gamma\delta}
\Biggr)\right.
\nonumber \\
&   &
\:\:\:\:\:\:\:\:\:\:
\left.
+
\cos\left( \vec{p}+\vec{q}\;\right)\cdot\hat{n}_l
\Biggl(
\bigl(\Sigma_l\bigr)^\dagger_{\rho\nu}\bigl(\Sigma_l\bigr)^\dagger_{\gamma\delta}
+
\bigl(\Sigma_l\bigr)_{\rho\nu}\bigl(\Sigma_l\bigr)_{\gamma\delta}
\Biggr)\right]  
\nonumber \\
&   &
-\frac{1}{N_{\text{c}}}\frac{K}{4} \sum_{\vec{p}} \sum_l \alpha_q
\left[ 2\alpha_p^2\, \Lambda^+_{\nu\gamma}(\vec{p}\,) - \delta_{\nu\gamma} \right]
\nonumber \\
&   &
\otimes\; \xi^\dagger_\rho(\vec{q}\,)\left[ N_{\text{c}}
\Biggl(
\bigl(\Sigma_l\bigr)_{\rho\nu}\bigl(\Sigma_l\bigr)^\dagger_{\gamma\delta}
+
\bigl(\Sigma_l\bigr)^\dagger_{\rho\nu}\bigl(\Sigma_l\bigr)_{\gamma\delta}
\Biggr)\right.
\nonumber \\
&   &
\:\:\:\:\:\:\:\:\:\:
\left.
+
\cos\left( \vec{p}+\vec{q}\;\right)\cdot\hat{n}_l
\Biggl(
\bigl(\Sigma_l\bigr)^\dagger_{\rho\nu}\bigl(\Sigma_l\bigr)^\dagger_{\gamma\delta}
+
\bigl(\Sigma_l\bigr)_{\rho\nu}\bigl(\Sigma_l\bigr)_{\gamma\delta}
\Biggr)\right] \Biggr\} \eta_\delta(-\vec{q}\,)
\nonumber \\
&   &
\:\:\:\:\:\:\:\:\:\:
\otimes B^\dagger_{\alpha,a}(\vec{q}\,)D^\dagger_{\alpha,a}(-\vec{q}\,)
|\;{\cal G}(0, \mu)\rangle
\label{eq:hoff2}
\end{eqnarray}
We see from Eq.~(\ref{eq:hoff2}) that the elementary excitations of the effective
Hamiltonian are color singlet (quasi) quark--anti--quark excitations coupled
to zero total three momentum. They correspond to the meson propagation on the lattice 
responsible for lifting the degeneracy of the ground state of strongly coupled QCD. 

With the use of the equation of motion for the $\Psi$ field Eq.~(\ref{eq:EOM1}), the equation 
of motion for the $\eta$ spinor Eq.~(\ref{eq:eometa}) and the orthonormality condition 
for the $\xi$ and $\eta$ spinors Eq.~(\ref{eq:norm2}), we can show that
\begin{eqnarray}
H_{\text{off}}|{\cal G}(0, \mu)\rangle
& = & 
\sum_{\vec{q}}
\Biggl\{ \alpha_q \xi^\dagger_\rho(\vec{q}\,)
\Bigl[ -\sum_l \text{sin}(\vec{q}\cdot\hat{n}_l)(\gamma_0\gamma_l)_{\rho\delta}
- M(\vec{q}\,)(\gamma_0)_{\rho\delta} + \mu_{\text{tot}}\delta_{\rho\delta} \Bigr]
\eta_\delta(-\vec{q}\,) \Biggr\} 
\nonumber \\
&   &
\:\:\:\:\:\:\:\:\:\:
\otimes B^\dagger_{\alpha,a}(\vec{q}\,)D^\dagger_{\alpha,a}(-\vec{q}\,)
|{\cal G}(0, \mu)\rangle
\nonumber \\
& = &
\sum_{\vec{q}} \Biggl\{
\alpha_q \xi^\dagger_\rho(\vec{q}\,)
\Bigl[\omega(\vec{q}\,) + \mu_{\text{tot}}\Bigr]\eta_\rho(-\vec{q}\,)
\Biggr\}
B^\dagger_{\alpha,a}(\vec{q}\,)D^\dagger_{\alpha,a}(-\vec{q}\,)
|{\cal G}(0, \mu)\rangle
\nonumber \\
& = &
0
\end{eqnarray}
Note that this result is valid for any dynamical quark mass and total chemical potential. 
Therefore in the $T \rightarrow 0$ limit our ansatz shown in Eq.~(\ref{eq:HAAGTMU}) exactly 
diagonalizes the effective Hamiltonian to second order in field operators for all densities.

Having diagonalized the second order Hamiltonian we can proceed to evaluate the vacuum 
energy density. Using Eq.~(\ref{eq:EOM1}) once more we find
\begin{eqnarray}
\frac{1}{V} \langle\, {\cal G}(0, \mu)\,| H_{\text{eff}} |\,{\cal G}(0, \mu)\, \rangle
& = &
- N_c N_f \sum_{\vec{p}} \Biggl[
\alpha_p^2 M_0 \text{Tr}\left( \Lambda^+(\vec{p}\,)\gamma_0\right)
+ 2\beta_p^2 \mu_0 \Biggr] 
\nonumber \\
&   &
-K \sum_{\vec{p}, \vec{q}}\sum_l \alpha_p^2 \alpha_q^2
\Lambda^+_{\nu\rho}(\vec{p}\,) \Lambda^+_{\delta\gamma}(\vec{q}\,)
\nonumber \\
&   &
\otimes\left[ \cos\left( \vec{p}-\vec{q}\;\right)\cdot\hat{n}_l
\Biggl(
\bigl(\Sigma_l\bigr)_{\rho\nu}\bigl(\Sigma_l\bigr)^\dagger_{\gamma\delta}
+
\bigl(\Sigma_l\bigr)^\dagger_{\rho\nu}\bigl(\Sigma_l\bigr)_{\gamma\delta}
\Biggr)\right.
\nonumber \\
&   &
\:\:\:\:\:\:\:\:\:\:
\left.
+
\cos\left( \vec{p}+\vec{q}\;\right)\cdot\hat{n}_l
\Biggl(
\bigl(\Sigma_l\bigr)^\dagger_{\rho\nu}\bigl(\Sigma_l\bigr)^\dagger_{\gamma\delta}
+
\bigl(\Sigma_l\bigr)_{\rho\nu}\bigl(\Sigma_l\bigr)_{\gamma\delta}
\Biggr)\right]
\nonumber \\
&   &
+ \frac{K}{2} \sum_{\vec{p}, \vec{q}}\sum_l 
\alpha_p^2 \left[ \alpha_q^2 \Lambda^+_{\nu\gamma}(\vec{q}\,) 
- \delta_{\nu\gamma} \right] \Lambda^+_{\delta\rho}(\vec{p}\,)
\nonumber\\
&   &
\otimes\left[ N_{\text{c}}
\Biggl(
\bigl(\Sigma_l\bigr)_{\rho\nu}\bigl(\Sigma_l\bigr)^\dagger_{\gamma\delta}
+
\bigl(\Sigma_l\bigr)^\dagger_{\rho\nu}\bigl(\Sigma_l\bigr)_{\gamma\delta}
\Biggr)\right.
\nonumber \\
&   &
\:\:\:\:\:\:\:\:\:\:
\left.
+
\cos\left( \vec{p}+\vec{q}\;\right)\cdot\hat{n}_l
\Biggl(
\bigl(\Sigma_l\bigr)^\dagger_{\rho\nu}\bigl(\Sigma_l\bigr)^\dagger_{\gamma\delta}
+
\bigl(\Sigma_l\bigr)_{\rho\nu}\bigl(\Sigma_l\bigr)_{\gamma\delta}
\Biggr)\right]
\nonumber\\
& = &
-2 N_c \sum_{\vec{p},\vec{q}} \Biggl\{
\alpha_p^2 \Biggl[ \frac{3}{2} K (1 + r^2) + \omega(\vec{p}\,)
+ \frac{M(\vec{p}\,)}{\omega(\vec{p}\,)} M_0
\nonumber\\
&   &
-\frac{1}{N_c} \frac{K}{2} (1 - r^2) 
\cos(\vec{p} + \vec{q}\,)\cdot\hat{n}_l - \mu_{\text{tot}} \Biggr]
+ (1 + \beta_p^2) \mu_0 \Biggr\}
\label{eq:VED}
\end{eqnarray}
For free space the difference of the vacuum energy densities in the Wigner--Weyl 
($M(\vec{q}\,) = 0$) and Nambu--Goldstone ($M(\vec{q}\,) \neq 0$) phases of the theory is 
positive
\begin{equation}
\Delta E = 
\frac{1}{V}\langle {\cal G}|H_{\rm eff}|{\cal G} \rangle |_{M(\vec{q}\,) = 0}
- \frac{1}{V}\langle {\cal G}|H_{\rm eff}|{\cal G} \rangle |_{M(\vec{q}\,) \neq 0}
> 0
\label{eq:VDENS}
\end{equation}
Numerically we find that Eq.~(\ref{eq:VDENS}) also holds for finite chemical potential. 
Therefore the true ground state of our interacting many body system is in the
phase with broken chiral symmetry.

\subsection{\label{Section4.3}Dynamical quark mass and $\mu_{\text{tot}}$}
We now derive the equations for the dynamical quark mass and the total chemical 
potential and solve them to determine our solution Eq.~(\ref{eq:HAAGTMU}) for each
density at zero temperature. To accomplish this we explicitly evaluate the right hand side of 
Eq.~(\ref{eq:EOM1}) to reveal its Dirac structure. The result may be cast in the 
following compact form
\begin{eqnarray}
\lefteqn{ \!\!\!\!\!\!\!\!\!\!\!\!\!\!\!
\Bigl[ \sum_l \sin(\vec{q}\cdot\hat{n}_l)(\gamma_0\gamma_l)_{\nu\delta}
+ M(\vec{q}\,)(\gamma_0)_{\nu\delta} - \mu_{\rm tot}\delta_{\nu\delta} \Bigr]
\bigl(\Psi_{a\alpha}\bigr)_\delta(t, \vec{q}\,) =
} \nonumber \\
& &
\;\;\;\;\;\;\;\;\;\;\;
\Bigl[ A(\vec{q}\,)(\gamma_0\gamma_l)_{\nu\delta} +
B(\vec{q}\,)(\gamma_0)_{\nu\delta} + C(\vec{q}\,)\delta_{\nu\delta} \Bigr]
\bigl(\Psi_{a\alpha}\bigr)_\delta(t, \vec{q}\,)
\label{eq:EOM}
\end{eqnarray}
The equations for $M(\vec{p}\,)$ and $\mu_{\rm tot}$ are obtained by equating the 
coefficents of the $\gamma_0$ operator and the Kronecker delta function, respectively. 

The gap equation determining $M(\vec{p}\,)$ is given by the coefficient $B(\vec{q}\,)$
\begin{eqnarray}
M(\vec{q}\,)
& = &
B(\vec{q}\,)
\nonumber \\
& = &
M_0 + \frac{3}{2}K(1 - r^2) \sum_{\vec{p}} \left( 1 - \beta_p^2 \right)
\frac{M(\vec{p}\,)}{\omega(\vec{p}\,)}
\nonumber \\
&   &
+ \frac{K}{N_{\text{c}}} \sum_{\vec{p},l} \left( 1 - \beta_p^2 \right)
\frac{M(\vec{p}\,)}{\omega(\vec{p}\,)}
\nonumber \\
&   &
\otimes
\Biggl\{
8r^2  \cos(\vec{p}\cdot\hat{n}_l) \cos(\vec{q}\cdot\hat{n}_l)
-\frac{1}{2}(1+r^2) \cos(\vec{p}+\vec{q}\, )\cdot\hat{n}_l
\Biggr\}
\label{eq:GAPEQ}
\end{eqnarray}
The structure of this gap equation is very similar to the one 
in free space ($\beta^2_p = 0$) found in \cite{umi00a}. The dynamical quark
mass is a constant to lowest order in $N_{\text{c}}$ but becomes momentum dependent
once $1/N_{\text{c}}$ correction is taken into account. 

Similarly, the total chemical potential is given by the coefficient $C(\vec{q}\,)$ 
\begin{eqnarray}
\mu_{\rm tot}
& = &
- C(\vec{q}\,) \nonumber \\
& = &
\mu_0 + \frac{1}{4} \frac{K}{N_{\text{c}}} \sum_l\sum_{\vec{p}} \beta_p^2
\Bigl[ 2N_{\text{c}} \left( 1+r^2 \right) - 2 \left( 1-r^2 \right)
{\rm cos}\left( \vec{p} + \vec{q}\, \right) \Bigr]
\label{eq:MUTOT}
\end{eqnarray} 
Thus $\mu_{\text{tot}}$ is a sum of the bare chemical potential $\mu_0$ and an interaction 
induced chemical potential which is proportional to the effective coupling constant $K$. 
Furthermore, the latter contribution to $\mu_{\text{tot}}$ is momentum dependent and this 
dependence is a $1/N_{\text{c}}$ correction just as in the case of the gap equation. It 
should be noted that the above shifting of the bare chemical potential by the interaction 
is not a new effect. For example, in the well--known and well--studied Nambu--Jona--Lasinio 
model \cite{nam61} at finite $T$ and $\mu$ the interaction induces a contribution to the 
total chemical potential which is proportional to the number density \cite{asa89,kle92}. 

The two equations Eqs.~(\ref{eq:GAPEQ}) and (\ref{eq:MUTOT}) are coupled and therefore 
solutions for $M$ and $\mu_{\text{tot}}$ must be found self--consistently for each value of the 
input parameter $\mu_0$. We shall solve the coupled equations to ${\cal O}(N_{\text{c}}^0)$ 
which is equivalent to invoking the mean field approximation. At this order in $N_{\text{c}}$ 
both the dynamical mass and the total chemical potential are momentum independent. It is 
also the same order in the $1/N_{\text{c}}$ expansion used to obtain results in all previous 
studies of strongly coupled lattice QCD. All results are presented using 
$M_0$ = 0 and $N_{\text{c}} = 3$.

%
%
\begin{figure}[tbp]
\begin{center}
\includegraphics[height=13cm,width=7cm,angle=-90]{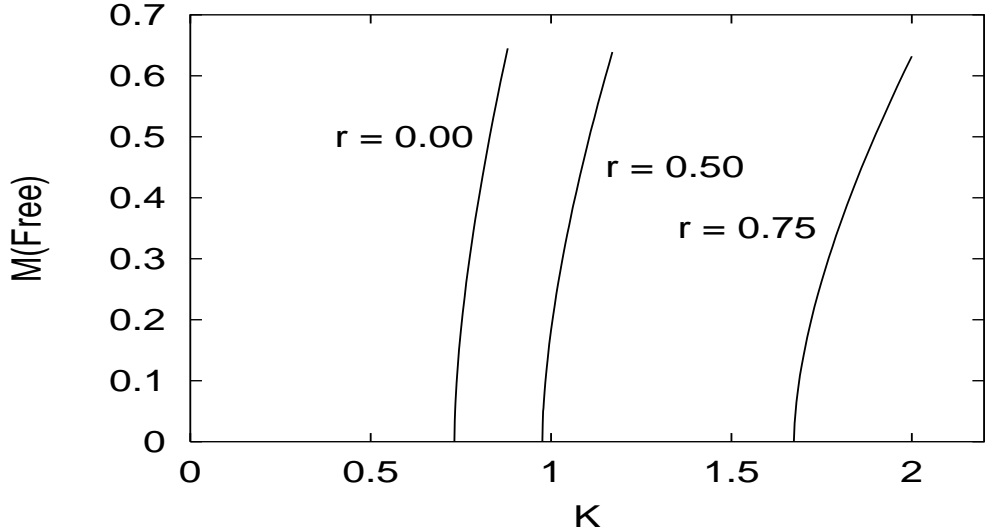} 
\end{center}
\caption{\label{fig1}Dynamical quark masses in free space $M(\text{Free})$ to 
${\cal O}(N_{\text{c}}^0)$ as functions of the effective coupling constant $K$ obtained 
using Wilson parameters $r$ = 0.00, 0.50 and 0.75. Critical coupling constants are 
$K_{\text{C}}$ = 0.732, 0.976 and 1.673 for $r$ = 0.00, 0.50 and 0.75, respectively. 
} 
\end{figure}
We first discuss the solutions to the gap equation in free space. In Figure~\ref{fig1} we show 
dynamical quark masses in free space $M(\text{Free})$ as functions of the coupling 
constant $K$ for Wilson parameters $r$ = 0.00, 0.50 and 0.75. The dynamical quark masses 
were obtained in a straightfoward manner by solving the free space gap equation
\begin{equation}
M(\text{Free}) = M_0 + \frac{3}{2}K(1 - r^2) \sum_{\vec{p}} 
\frac{M(\text{Free})}{\omega(\vec{p}\,)}
\label{eq:FREEGAPEQ}
\end{equation}
The figure shows that for each value of $r$ there exists a critical coupling constant 
$K_{\text{C}} > 0$ above which the theory is in the broken symmetry phase. This is 
also true for the $r = 0$ case corresponding to the use of Kogut--Susskind fermions. In 
this case the symmetry breaking takes place only for $K \geq 0.732$. 

The dependence of the dynamical mass, and consequently of the chiral condensate through 
Eq.~(\ref{eq:cond}), on the coupling constant is qualitatively different from the 
results obtained previously using the {\em same} effective Hamiltonian \cite{smi80,ley86}. 
In both \cite{smi80} and \cite{ley86} $q\bar{q}$ pair condensation occurs for {\em any} 
value of $K > 0$. We find that the attraction between a quark and an anti--quark 
must be sufficiently large enough for a $q\bar{q}$ pair to condensate in the vacuum. Thus 
our approach provides a mechanism for chiral symmetry breaking which other approaches do not.
In addition, our results are consistent with the works by Finger and Mandula \cite{fin82} 
and by Amer, Le Yaouanc, Oliver, Pene and Raynal \cite{ame83} who have shown that in QCD 
in the Coulomb gauge $q\bar{q}$ condensation takes place only above a critical coupling 
constant.

%
%
\begin{figure}[tbp]
\begin{center}
\includegraphics[height=13cm,width=7cm,angle=-90]{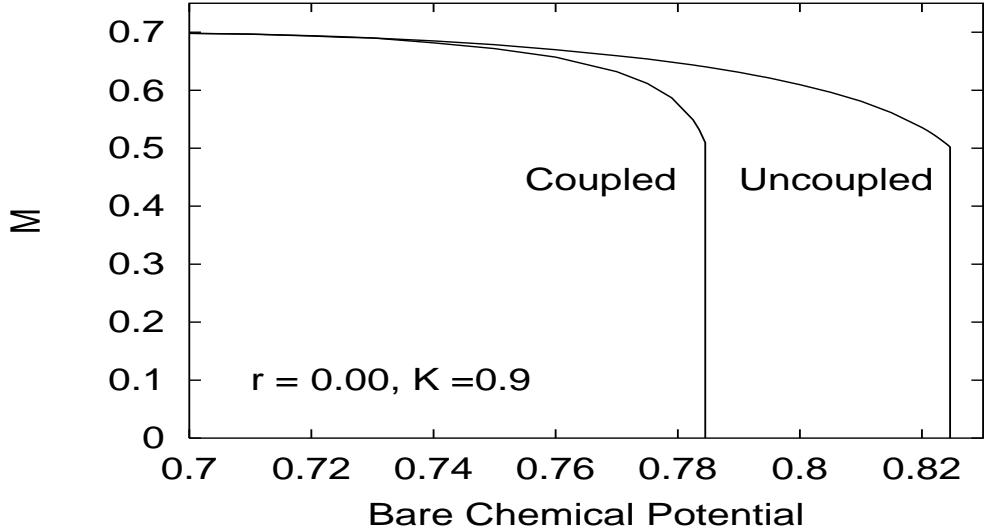} 
\end{center}
\caption{\label{fig2}Dynamical quark masses $M$ as functions of the bare chemical 
potential $\mu_0$ to lowest order in $N_{\text{c}}$. The dynamical mass labelled 
``Uncoupled'' was obtained by simply solving Eq.~(\ref{eq:GAPEQ}) with 
$\mu_{\text{tot}} = \mu_0$ and exhibits a first order phase transition with a critical 
chemical potential of $(\mu_0)_{\text{C}} = 0.825$. The result labelled ``Coupled'' was 
obtained by solving the coupled equations Eqs.~(\ref{eq:GAPEQ}) and (\ref{eq:MUTOT}) 
self--consistently. A first order phase transition also takes place, but now the value 
of $(\mu_0)_{\text{C}}$ is 0.785. The Wilson parameter and the coupling constant are set 
to $r = 0.0$ and $K = 0.9$, respectively.} 
\end{figure}

\begin{figure}[tbp]
\begin{center}
\includegraphics[height=13cm,width=7cm,angle=-90]{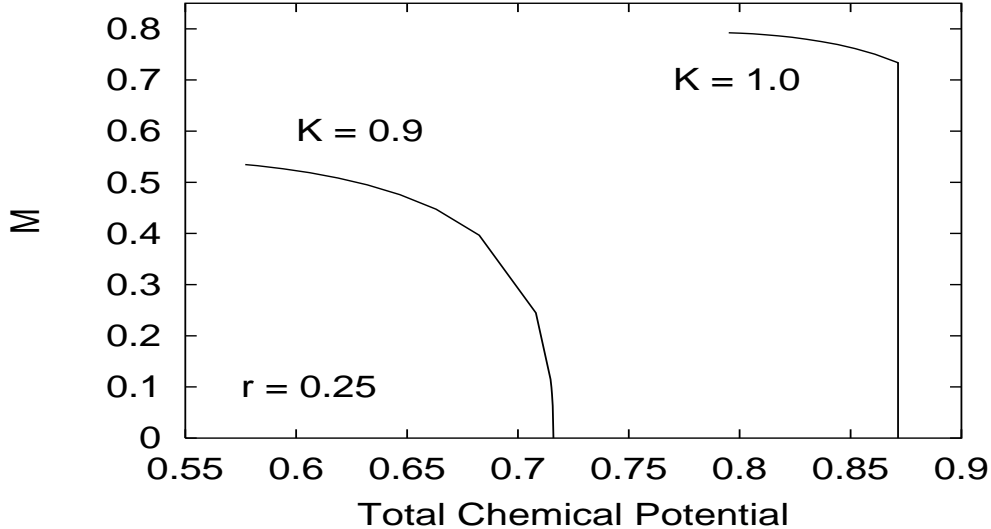} 
\end{center}
\caption{\label{fig3}Dynamical quark mass $M$ as a function of total chemical potential 
$\mu_{\text{tot}}$ for two values of the effective coupling constant $K$. These results were 
obtained by solving Eqs.~(\ref{eq:GAPEQ}) and (\ref{eq:MUTOT}) self--consistently to lowest 
order in $N_{\text{c}}$ using $r = 0.25$. There is a second 
order chiral phase transition when the effective coupling constant $K$ is 0.9 with a 
critical chemical potential of $(\mu_{\text{tot}})_{\text{C}} \approx 0.716$. The order of 
the phase transition becomes first order with $(\mu_{\text{tot}})_{\text{C}} \approx 0.871$ 
when $K$ is increased to 1.0.}
\end{figure}

Examples of finite $\mu$ solutions to the coupled equations Eqs.~(\ref{eq:GAPEQ}) and 
(\ref{eq:MUTOT}) to lowest order in $N_{\text{c}}$ are shown in Figures~\ref{fig2} and 
\ref{fig3}. In 
Figure~\ref{fig2} we show dynamical masses as functions of the {\em bare} chemical 
potential $\mu_0$ to highlight the importance of solving the coupled equations consistently.
The figure shows the dynamical quark mass obtained by solving only Eqs.~(\ref{eq:GAPEQ}) with 
$\mu_{\text{tot}} = \mu_0$ as well the mass obtained by solving the coupled equations 
consistently. Using $r = 0$ and $K = 0.9$ a first order phase transition is observed in 
both cases, but the values of the critical $\mu_0$ are 0.825 when $\mu_{\text{tot}} = \mu_0$ 
and 0.785 when the two equations are solved self--consistently. Therefore the critical 
chemical potential will be overestimated if interaction induced corrections to the bare 
chemical potential are ignored.

In Figure~\ref{fig3} we present the dynamical mass as a function of the {\em total} chemical 
potential $\mu_{\text{tot}}$ for two values of $K$ obtained with $r = 0.25$. From the figure 
we see that the phase transition can be either first or second order depending on the value 
of the coupling constant. When $K$ = 0.9 we find a second order phase transition with a 
critical chemical potential of $(\mu_{\text{tot}})_{\text{C}} \approx 0.716$, while if the 
coupling constant is increased to $K$ = 1.0 the phase transition becomes first order with a 
larger critical chemical potential of $(\mu_{\text{tot}})_{\text{C}} \approx 0.871$. This 
increase in the critical chemical potential with $K$ has also been observed in \cite{bil92}.
Furthermore, we find that when $K = 0.9$ lattice saturation sets in above 
$(\mu_{\text{tot}})_{\text{C}}$ at around $\mu_{\text{tot}} \approx 0.898$ while 
this effect takes place immediately above $(\mu_{\text{tot}})_{\text{C}}$ for $K = 1.0$. 
These values of chemical potentials are smaller than the energy $E=1/K$ required to 
excite one color electric flux link as given in Eq.~(\ref{eq:ELINK}). Therefore with a 
reasonable set of parameters it is possible to extend Smit's effective Hamiltonian to finite 
density as was first pointed out in \cite{ley88}. 

%
%
%

%
\begin{figure}[tbp]
\begin{center}
\includegraphics[height=13cm,width=7cm,angle=-90]{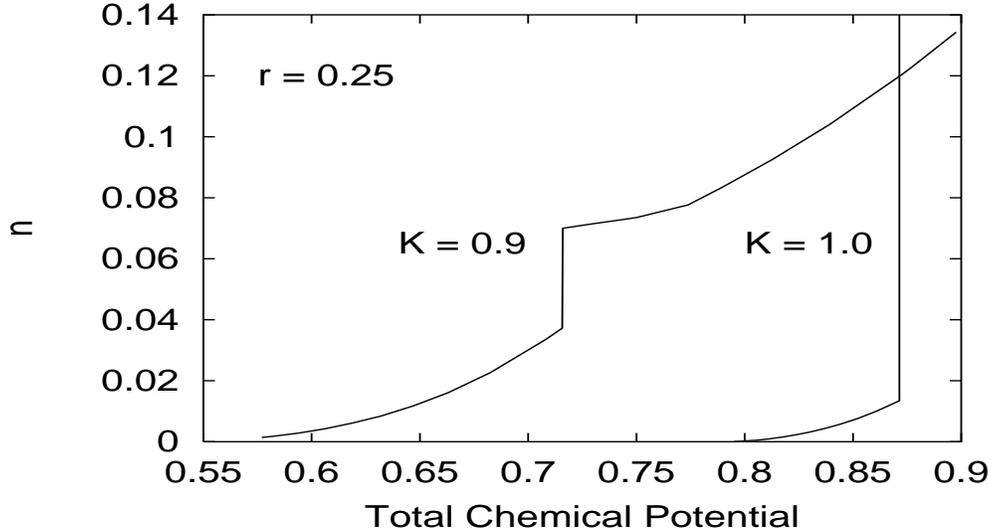} 
\end{center}
\caption{\label{fig4}Quark number density $n$ as a function of total chemical potential 
$\mu_{\text{tot}}$ for two values of effective coupling constant $K$ with $r = 0.25$. 
When $K = 0.9$ there is a jump in the number density at the phase transition point at 
$(\mu_{\text{tot}})_{\text{C}} \approx 0.716$ from 
$n \approx 0.037$ to $n \approx 0.070$, while for $K = 1.0$ $n$ becomes 
unity immediately above the critical chemical potential of 
$(\mu_{\text{tot}})_{\text{C}} \approx 0.871$ due to lattice saturation.}
\end{figure}

Having solved the self--consistency equations for the dynamical quark mass and the total 
chemical potential to lowest order in $N_{\text{c}}$ we have constructed a mean field 
solution for the quark field appearing in the effective Hamiltonian Eq.~(\ref{eq:Heffp}). 
In Figure~\ref{fig4} we show the quark number density obtained with this 
solution as a function of $\mu_{\text{tot}}$ for $K$ = 0.9 and 1.0. In both cases the number 
density is a monotonically increasing function of $\mu_{\rm tot}$ in the broken symmetry 
phase. When $K = 0.9$ there is a jump in the number density at the phase transition point at
$(\mu_{\text{tot}})_{\text{C}} \approx 0.716$ from $n \approx 0.037$ to $n \approx 0.070$. 
Beyond this point the number density continues to increase monotonically until when the lattice 
saturation sets in at $\mu_{\text{tot}} \approx 0.898$. This behaviour of the number density 
is qualitatively the same as the one obtained numerically using the Monomer--Dimer--Polymer 
algorithm as can be seen from a comparison with 
Figure~5 of \cite{kar89}. For $K = 1.0$ the lattice saturation 
takes place at the phase transition point at $(\mu_{\text{tot}})_{\text{C}} \approx 0.871$ and
beyond this point the number density remains a constant at $n = 1$. Noting that the number 
density at the phase transition point is $n \approx 0.013$, the number density for $K = 1.0$ 
may be approximated by a 
Heaviside function of the form $n = \theta( \mu_{\text{tot}} - (\mu_{\text{tot}})_{\text{C}})$.
This is exactly the result obtained in Eq.~(2.47) of \cite{gre00} where a 
different effective Hamiltonian was used to study strongly coupled lattice QCD at finite 
density. Furthermore the functional form of the number density found in \cite{gre00} is 
independent of the strength of the interaction. Therefore the results presented in 
\cite{gre00} for $\mu \geq \mu_{\text{C}}$ represent nothing but those obtained in the 
lattice saturation limit.

\section{\label{Section5}Equation of State}
%
%
%
\begin{figure}[tbp]
\begin{center}
\includegraphics[height=13cm,width=7cm,angle=-90]{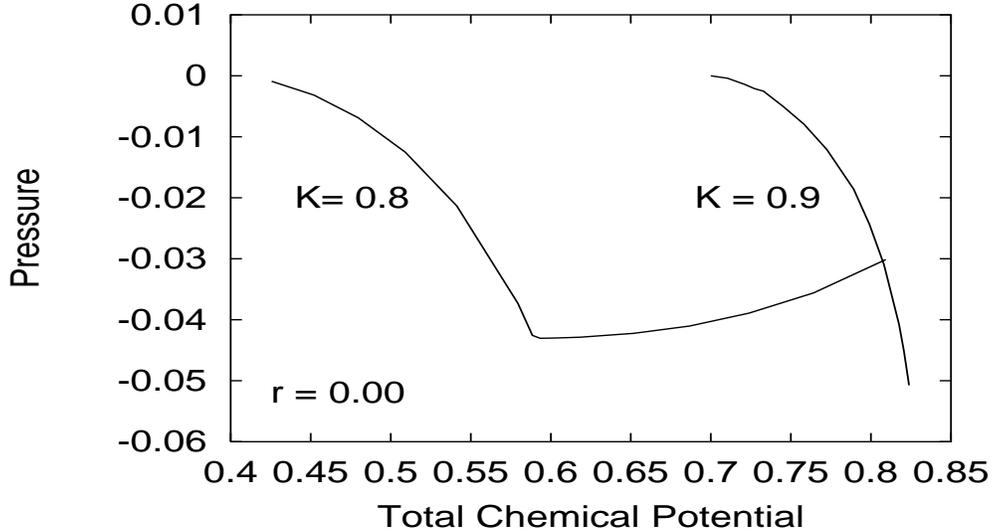} 
\end{center}
\caption{\label{fig5}Pressure as a function of total chemical potential 
$\mu_{\text{tot}}$ obtained using Kogut--Susskind fermions ($r$ = 0.0) with $K$ = 0.8 and 0.9.}
\end{figure}
\begin{figure}[tbp]
\begin{center}
\includegraphics[height=13cm,width=7cm,angle=-90]{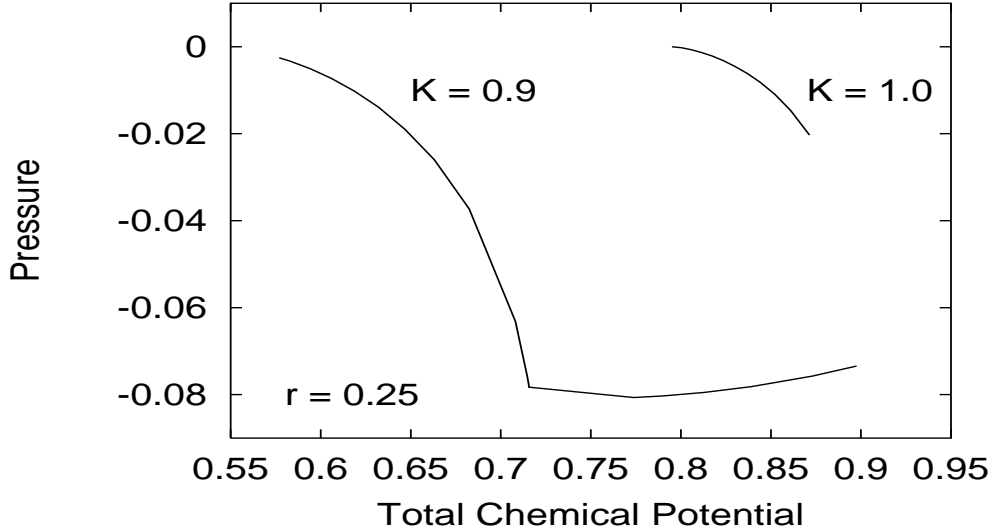} 
\end{center}
\caption{\label{fig6}Pressure as a function of total chemical potential 
$\mu_{\text{tot}}$ obtained using Wilson fermions ($r$ = 0.25) with $K$ = 0.9 and 1.0.}
\end{figure}

We are now in a position to determine the equation of state by numerically evaluating the 
thermodynamic potential density using the mean field solution determined above in the vacuum 
energy density Eq.~(\ref{eq:VED}). In Figure~\ref{fig5} we plot pressure as a function 
of $\mu_{\text{tot}}$ for $K$ = 0.8 and 0.9. The value 
of the Wilson parameter is $r = 0.0$ so that the results have been obtained using 
Kogut--Susskind fermions. For both 
values of $K$ we find that the pressure of the quark Fermi sea is negative and monotonically 
decreasing in the broken symmetry phase. For $K$ = 0.8 the pressure remains negative but 
increases in the symmetry restored phase, at least until the lattice saturation point, and has 
a cusp where the two phases meet. Unfortunately, for $K = 0.9$ we can not make a definite 
quantitative statement concerning the behaviour of the pressure in the symmetry restored 
phase due to lattice saturation, except to mention that there is a discontinuity when going 
from one phase to another. We find qualitatively similar results when Wilson fermions are 
used to calculate the pressure as shown in Figure~\ref{fig6}. The parameter used in this 
figure are $r$ = 0.25 and $K$ = 0.9 and 1.0. We may therefore 
conclude, at least at the mean field level, that up to and beyond the chiral symmetry 
restoration point the quark Fermi sea can have negative pressure and therefore can be 
mechanically unstable with an imaginary speed of sound.

%
%

Our conclusion regarding the (strongly coupled) quark matter stability at finite density 
is consistent with similar studies using effective continuum models of QCD. In the 
Nambu--Jona--Lasinio model \cite{bub96} and the instanton induced 't~Hooft interaction model 
\cite{alf98}, mean
field calculations show that cold and dense quark matter may be unstable in the phase with 
spontaneously broken chiral symmetry, but can become stable in the symmetry restored phase 
at high enough density. In particular, the result for the pressure obtained in \cite{alf98} 
is qualitatively the same as the one shown in Figures~\ref{fig5} and \ref{fig6} as can be
seen by comparing the figures with Figure~1 of \cite{alf98}. The possibility of unstable 
quark mattter lead the authors of \cite{bub96} and \cite{alf98} to speculate the 
formation of nucleon droplets, reminiscent of the MIT bag model, in the broken symmetry phase. 
We shall not indulge on such a speculation here since we are working in an artificial strong 
coupling regime. Nevertheless, our results concerning the negative pressure is certainly 
verifiable in future lattice simulations of finite density QCD at strong coupling.

\section{\label{Section6}Conclusion and Outlook}

In this work we studied the equation of state of two flavored Hamiltonian lattice QCD in the 
strong coupling limit at finite density using both Kogut--Susskind and Wilson fermions. 
Starting from an effective lattice Hamiltonian for the ground state of the strongly coupled 
QCD, we constructed a mean field solution which exactly diagonalizes the Hamiltonian to 
second order in field operators for all densities. This solution obeys the free lattice 
Dirac equation with a dynamical quark mass and total chemical potential which are determined 
by solving a coupled set of equations obtained from the equation of motion. From the gap
equation determining the dynamical quark mass we find that at the mean field level the order 
of the chiral phase transition can be either first or second order depending on the values 
of input parameters. 

The equation of state was obtained by evaluating the thermodynamic potential density from 
the vacuum energy density using our solution. We find that the pressure of the strongly 
interacting many body system may be negative in the broken symmetry phase indicating the 
mechanical instability of our quark Fermi sea. There are indications of this instability 
beyond the phase transition point although no definite conclusions could be reached for 
very high densities due to lattice saturation. Nevertheless this behaviour of the pressure 
was found both for the case of Kogut--Susskind and Wilson fermions and seems, at least at
the mean field level, to be robust. In addition, our result concerning negative pressure 
is in qualitative agreement with studies using continuum effective QCD models, and therefore 
should certainly be verified by future lattice simulations of strongly coupled QCD at 
finite density.

To include temperature into our formalism we simply repeat our calculations
using the ansatz given in Eq.~(\ref{eq:HAAGTMU}) at non--zero $T$. Preliminary
calculations indicated that, in addition to particle--anti--particle excitations,
the elementary excitations would now involve particle--hole, anti--particle--anti--hole
and hole--anti--hole excitations. Because of these additional types of
excitations our ansatz would no longer be able
to exactly diagonalize the second order Hamiltonian. In fact, a simple
exercise would show that at finite $T$ even the free lattice Dirac
Hamiltonian Eq.~(\ref{eq:H0}) is not diagonal due to particle--hole and
anti--particle--anti--hole excitations.

We now turn our attention to the possibility of studying the nature of the 
confinement--deconfinement phase transition. Our solution presented in this work 
is non--confining and therefore it would be hopeless to use it to study this 
important phase transition. What is lacking in our formalism is the
description of bound states. However, our solution presented here is by no means 
unique or complete and it can be systematically improved to include all the bound
states allowed by the effective Hamiltonian. This is
accomplished by interpreting our solution within the context of the N--quantum 
approach (NQA) to quantum field theory \cite{gre65,gre94}.

NQA is a method to solve field equations of motion by expanding the interacting Heisenberg 
fields in terms of asymptotic fields obeying the free field equations of motion. 
Here the on--shell masses can but need not equal the physical masses of the fields. This 
expansion 
is known as the Haag expansion \cite{haa55} and our ansatz presented here is nothing 
but the first term in this expansion. Note that because we are working in
the Hamiltonian (Kogut--Susskind) formulation of lattice field theory the
time variable is continuous and therefore we can introduce and work with the 
concept of asymptotic fields. The second order terms in the Haag expansion 
would consist of a product of fermionic quark fields and bosonic {\em elementary} 
color singlet $\bar{q}q$ bound state fields. The coefficient of each of the second order 
terms are interpreted as creation amplitudes for the bound states and are known as Haag 
amplitudes.

Supressing color and flavor indices for simplicity, our extended ansatz for
the Heisenberg quark field $\Psi$ to second order in the Haag
expansion in free space will have the following structure
\begin{eqnarray}
\Psi_\nu^{\text{Free}}(\vec{q}\,)
& = &
\Psi^0_\nu(\vec{q}\,) 
+ 
\sum_i
\int d^3k d^3b\; \delta^3(\vec{q} + \vec{k} - \vec{b}\,)
f^{(i)}_{\nu\rho}(\vec{k}, \vec{b}\,)
:\Psi^0_\rho(-\vec{k}\,) B^0_{(i)}(\vec{b}\,): 
\label{eqn:haag} 
\end{eqnarray}
where the symbol $:\hspace{0.5cm}:$ denotes normal ordering. In Eq.~(\ref{eqn:haag}),
$B$ is the elementary bosonic field while the superscript $0$ indicates that the fields
obey their corresponding free field equations of motion. The Haag amplitudes are denoted
by $f^{(i)}_{\nu\rho}$ with the sum over the index $i$ running through all the possible 
bound states allowed by the Hamiltonian. These states are the color singlet $q\bar{q}$ 
elementary excitations identified in this work. The basic idea of NQA is to use the 
field equations of motion and derive integral equations for the Haag amplitudes and 
solve them to obatin a solution to the equation of motion. 

In order to solve for the Haag amplitudes it is necessary to calculate the mass and
the coupling constant for each of the bound states. This has been accomplished
successfully at finite $T$ and $\mu$ for the two flavored `t~Hooft interaction model 
\cite{umi00b}. In addition to bound state masses and coupling constants it is also
possible to determine the widths of these states. This quantity is the key to studying
the confinement--deconfinement phase transition within our formalism. In the confined 
phase the bound states will have vanishing widths while in the deconfined phase we expect 
to see unbound resonant states with finite widths. Hence we propose to use the
widths of the $q\bar{q}$ states as an order parameter to study the nature of the 
deconfinement phase transition within strong coupling QCD.

\begin{acknowledgments}
Part of this work was completed while I was a Junior Visiting Scientist at
ECT$^*$ in Trento, Italy. I thank the Center for its hospitality and generous 
support.
\end{acknowledgments}
\appendix*
\section{\label{Appendix}Properties of free Wilson fermions}
In this Appendix we present the properties of free Wilson fermions on the lattice 
in the Hamiltonian formulation \cite{kog75}. The free lattice Dirac Hamiltonian is given by
\begin{eqnarray}
H^0
& = &
\frac{1}{2i} \sum_{\vec{x},l}  
\left[ \Psi^\dagger(\vec{x}\,) \gamma_0\gamma_l \Psi(\vec{x}+\hat{n}_l)
- \Psi^\dagger(\vec{x}+\hat{n}_l) \gamma_0\gamma_l \Psi(\vec{x}\,) \right]
+ M \sum_{\vec{x}} \Psi^\dagger(\vec{x}\,) \gamma_0 \Psi(\vec{x}\,)
\nonumber\\
&   &
- \frac{r}{2} \sum_{\vec{x},l}
\left[ \Psi^\dagger(\vec{x}\,) \gamma_0 \Psi(\vec{x}+\hat{n}_l)
+ \Psi^\dagger(\vec{x}+\hat{n}_l) \gamma_0 \Psi(\vec{x}\,) \right]
\end{eqnarray}
where the third term is the Wilson term. For $r = 0$ there is 
an eightfold fermion multiplicity which is removed when $r \neq 0$. 
At each lattice site the free Dirac field in configuration space is given by
\begin{equation}
\Psi_\nu(t,\vec{x}\,) = 
\sum_{\vec{p}} \left[ b(\vec{p}\,)\xi_\nu(\vec{p}\,)
e^{-i(\omega(\vec{p}\,) t - \vec{p}\cdot\vec{x})}
+ d^\dagger(\vec{p}\,)\eta_\nu(\vec{p}\,)
e^{i(\omega(\vec{p}\,) t - \vec{p}\cdot\vec{x})} \right]
\end{equation}
with $\nu$ denoting the Dirac index. The excitation energy $\omega(\vec{p}\,)$
will be determined shortly. The annihilation operators $b$ and $d$
annihilate the non--interacting vacuum state $|\,0\,\rangle$. For our purpose 
it is not necessary to know the structure of the spinors $\xi$ and $\eta$. 

The only assumption that we shall make is that the creation and annihilation 
operators obey the free fermion anti--commutation relations
\begin{equation}
\left[b^\dagger(\vec{p}\,), b(\vec{q}\,)\right]_+
=
\left[d^\dagger(\vec{p}\,), d(\vec{q}\,)\right]_+
=
\delta_{\vec{p},\vec{q}}
\label{eq:antic}
\end{equation}
Using this assumption we can recover the anti--commutation relations for 
the field operators
\begin{equation}
\left[\Psi_\rho(t,\vec{x}\,), \Psi^\dagger_\nu(t,\vec{y}\,)\right]_+
=
\delta_{\vec{x},\vec{y}}\;\delta_{\rho\nu}
\end{equation}
provided that $\xi$ and $\eta$ satisfy the relation
\begin{equation}
\xi_\rho(\vec{p}\,) \xi^\dagger_\nu(\vec{p}\,) +
\eta_\rho(-\vec{p}\,) \eta^\dagger_\nu(-\vec{p}\,)
=
\delta_{\rho\nu}
\label{eq:compl}
\end{equation}
We normalize the spinors by demanding that the number density is given by 
\begin{equation}
{\cal N} 
= 
\sum_{\vec{x}} :\Psi^\dagger(t,\vec{x}\,)\Psi(t,\vec{x}\,): 
\; = 
2 \sum_{\vec{p}}
\left( b^\dagger(\vec{p}\,) b(\vec{p}\,)
- d^\dagger(\vec{p}\,) d(\vec{p}\,)\right)
\label{eq:ndensity}
\end{equation}
where the symbol :    : denotes normal ordering with respect to $|\,0\,\rangle$ and the 
factor of 2 accounts for the spin degrees of freedom. Eq.~(\ref{eq:ndensity}) fixes 
the normalizations of $\xi$ and $\eta$ to be  
\begin{eqnarray}
\xi_\nu(\vec{p}\,) \xi^\dagger_\nu(\vec{p}\,) 
& = & \eta_\nu(\vec{p}\,) \eta^\dagger_\nu(\vec{p}\,)
 =  2 
\label{eq:norm1}\\
\xi^\dagger_\nu(\vec{p}\,) \eta_\nu(-\vec{p}\,)
& = &
\eta^\dagger_\nu(\vec{p}\,) \xi_\nu(-\vec{p}\,)  
 =  0 
\label{eq:norm2}
\end{eqnarray}
which are consistent with Eq.~(\ref{eq:compl}).

In momentum space the charge conjugaton symmetric form of $H^0$ is 
\begin{equation}
H^0 = 
\frac{1}{2} \sum_{\vec{p}} 
\Bigl( - \sum_{l} \sin(\vec{p}\cdot\hat{n}_l) \gamma_0\gamma_l
+ M(\vec{p}\,) \gamma_0 \Bigr)_{\rho\nu}
\biggl[ \Psi^\dagger_\rho(t,\vec{p}\,), \Psi_\nu(t,-\vec{p}\,) \biggr]_-
\label{eq:h0}
\end{equation}
where the momentum dependent mass term is given by 
\begin{equation}
M(\vec{p}\,) \equiv M - r \sum_l \cos(\vec{p}\cdot\hat{n}_l)
\label{eq:massp}
\end{equation}
The free Dirac field now becomes 
\begin{equation}
\Psi_\nu(t,\vec{p}\,) = 
b(\vec{p}\,)\xi_\nu(\vec{p}\,)e^{-i\omega(\vec{p}\,) t}
+ d^\dagger(-\vec{p}\,)\eta_\nu(-\vec{p}\,)e^{+i\omega(\vec{p}\,) t}
\label{eq:psipA} 
\end{equation}
which is used to derive the equation of motion corresponding to Eq.~(\ref{eq:h0})
\begin{eqnarray}
i \dot{\Psi}(t,\vec{p}\,)
& = &
:\left[\Psi(t,\vec{p}\,), H^0\right]_-:
\\
& = & 
\biggl( \sum_{l} \sin(\vec{p}\cdot\hat{n}_l) \gamma_0\gamma_l
+ M(\vec{p}\,) \gamma_0 \biggr) \Psi(t,\vec{p}\,)
\label{eq:eomh0}
\end{eqnarray}
From Eq.~(\ref{eq:eomh0}) one obtains the excitation energy
\begin{equation}
\omega(\vec{p}\,) =
\biggl( \sum_{l} \sin^2(\vec{p}\cdot\hat{n}_l) 
+ M^2(\vec{p}\,)\biggr)^{1/2}
\label{eq:EXENA}
\end{equation}
and the equations of motion for the $\xi$ and $\eta$ spinors
\begin{eqnarray}
\omega(\vec{p}\,)\; \xi(\vec{p}\,)
& = & 
\biggl( \sum_{l} \sin(\vec{p}\cdot\hat{n}_l) \gamma_0\gamma_l
+ M(\vec{p}\,) \gamma_0 \biggr) \xi(\vec{p}\,)
\label{eq:eomxi} \\
\omega(\vec{p}\,)\; \eta(-\vec{p}\,)
& = & 
- \biggl( \sum_{l} \sin(\vec{p}\cdot\hat{n}_l) \gamma_0\gamma_l
+ M(\vec{p}\,) \gamma_0 \biggr) \eta(-\vec{p}\,)
\label{eq:eometa}
\end{eqnarray}

When $r=0$ these equations of motion are relativistic near the eight corners
of the Brillouin zone denoted by
$\vec{\pi}_0 = (0, 0, 0)$,
$\vec{\pi}_{\text{x}} = (\pi, 0, 0)$,
$\vec{\pi}_{\text{y}} = (0, \pi, 0)$,
$\vec{\pi}_{\text{z}} = (0, 0, \pi)$,
$\vec{\pi}_{\text{xy}} = (\pi, \pi, 0)$,
$\vec{\pi}_{\text{xz}} = (\pi, 0, \pi)$,
$\vec{\pi}_{\text{yz}} = (0, \pi, \pi)$ and
$\vec{\pi}_{\text{xyz}} = (\pi, \pi, \pi)$. The excitation energies near these values
of momenta are equal which corresponds to the eightfold multiplicity mentioned
above. This degeneracy is lifted when $r \neq 0$ due to the momentum
dependent mass term Eq.~(\ref{eq:massp}). Using the equations of motion for $\xi$
and $\eta$ it is a simple excercise to show that the off--diagonal Hamiltonian 
vanishes and that the vacuum energy is given
by 
\begin{equation}
\langle\, 0\,|H^0|\,0\, \rangle = -2 V\sum_{\vec{p}} \omega(\vec{p}\,)
\end{equation}
Finally, we construct positive and negative energy projection 
operators $\Lambda^+(\vec{p}\,)$ and $\Lambda^-(\vec{p}\,)$ as follows
\begin{subequations}
\label{eq:proj}
\begin{eqnarray}
\Lambda^+(\vec{p}\,) 
& \equiv & 
\xi(\vec{p}\,)\otimes\xi^\dagger(\vec{p}\,) =
\frac{1}{2} \left[ 1 + 
\frac{1}{\omega(\vec{p}\,)} \sum_{l} \sin(\vec{p}\cdot\hat{n}_l) \gamma_0\gamma_l
+ \frac{M(\vec{p}\,)}{\omega(\vec{p}\,)} \gamma_0 \right]
\label{eq:projp} \\
\Lambda^-(\vec{p}\,)
& \equiv &
\eta(-\vec{p}\,)\otimes\eta^\dagger(-\vec{p}\,) =
\frac{1}{2} \left[ 1 -
\frac{1}{\omega(\vec{p}\,)} \sum_l \sin(\vec{p}\cdot\hat{n}_l) \gamma_0\gamma_l
- \frac{M(\vec{p}\,)}{\omega(\vec{p}\,)} \gamma_0 \right]
\label{eq:projn} 
\end{eqnarray}
\end{subequations}
Note that the projection operators obey the condition
\begin{equation}
\left[ \Lambda^+ (\vec{p}\,) + \Lambda^-(\vec{p}\,)\right]_{\rho\nu}
= \delta_{\rho\nu}
\end{equation}
as is required by Eq.~(\ref{eq:compl}). 
\bibliography{StrongQCD}
\end{document}